\documentstyle[graphicx]{mn}

\newif\ifAMStwofonts
\AMStwofontstrue

\def\2M{{2MASS 234449+1221}}

\def\xmm{{\it XMM-Newton}}
\def\chandra{{\it Chandra}}

\def\et{{et al.\ }}

\newcommand{\ls}{\mathrel{\hbox{\rlap{\hbox{\lower4pt\hbox{$\sim$}}}\hbox{$<$}}}}
\newcommand{\gs}{\mathrel{\hbox{\rlap{\hbox{\lower4pt\hbox{$\sim$}}}\hbox{$>$}}}}

\def\arcs{{\hbox{$^{\prime\prime}$}}}

\def\H0{{\rm ~km~s^{-1}~Mpc^{-1}}}

\def\et{{et al.}}

\title[\xmm\ observation of \2M]
        {X-ray absorption and re-emission from an ionised outflow in the Type 1 QSO \2M}
\author[K.A.Pounds \et]
        {K.A.Pounds$^{1}$,
         B.J.Wilkes$^{2}$
	 and K.L.Page$^{1}$ \\
$^1$ Department of Physics and Astronomy, University of Leicester,
Leicester, LE1 7RH, UK\\
$^2$ Harvard/Smithsonian Center for Astrophysics, Cambridge, MA 02138, USA\\}

\date{Accepted ; Submitted ; Revised }
\pagerange{\pageref{firstpage}--\pageref{lastpage}}
\pubyear{2004}
\begin{document}
\maketitle
\label{firstpage}

\begin{abstract}
We report on the analysis of a short \xmm\ observation of the reddened Type 1 QSO \2M\ first identified in the Two Micron All-Sky 
Survey. The underlying X-ray continuum is found to be typical of a broad-line active galaxy, with photon index $\Gamma$$\sim$1.9.
Low energy absorption can be modelled by a column N$_{H}$$\sim$$10^{22}$ cm$^{-2}$ of moderately ionised gas or a smaller column of
cold gas. Addition of a soft X-ray emission component significantly improves the fit in both cases. With the assumption that the
soft X-ray flux represents emission from gas photoionised by the incident X-ray continuum, a comparison of the absorbed and emitted
luminosities indicates a covering factor of $\sim$8-17\%. The unusual opportunity to simultaneously observe and quantify ionised
absorption {\it and} emission in \2M\ is due to the relatively large opacity (for a Type 1 AGN) of the absorbing gas, which
depresses the normally strong continuum below $\sim$1 keV. A comparison of the soft X-ray emission of \2M\ with that of other Type 1
and Type 2 AGN suggests the existence of an inner turbulent extension to ionised outflows, not detected in current high resolution
X-ray spectra. 
\end{abstract}


\section{Introduction}

X-ray spectra of Type 1 AGN are typically dominated by a power law continuum of photon index $\Gamma$$\sim$1.8-2 (Nandra and Pounds
1994), with an ionised outflow often imprinting an absorption line spectrum in the soft X-ray band (e.g. Kaspi \et\ 2002, Steenbrugge
\et\ 2003). Type 2 AGN, in contrast, are usually heavily absorbed by cold gas (perhaps in the putative torus; Antonucci 1993), with
soft X-ray lines showing up in emission. While it is a reasonable assumption that the soft X-ray emission and absorption come from
the same ionised outflow (Kinkhabwala \et\ 2002) both components are rarely seen in the same spectrum, since the continuum flux is
generally dominant in Type 1 AGN. A few cases have been found, however, where a Type 1 AGN in a low flux state exhibits a
dominantly emission line spectrum (e.g. Turner \et\ 2003, Pounds \et\ 2004b).  

The Two Micron All-Sky Survey (2MASS) has revealed many highly reddened active galaxies (AGN) whose number density rivals that of
optically selected AGN. Spectroscopic follow-up of red candidates reveals $\sim$75 percent are previously unidentified emission-line
AGN, with $\sim$80 percent of those showing the broad optical emission lines of Type 1 Seyfert galaxies and QSOs (Cutri \et\ 
2001). These objects often have unusually high optical polarization levels, with $\sim$10 percent showing $P>3$ percent indicating a
significant contribution from scattered light (Smith \et\ 2002) and suggesting substantial obscuration toward the nuclear energy
source. \chandra\ observations of a sample of 2MASS AGN found them to be X-ray weak with generally flat (hard) spectra (Wilkes \et\
2002).

\xmm\ spectra were recently obtained for a subset of five 2MASS AGN in an attempt resolve the effects of absorption from an
{\it intrinsically} flat power law spectrum. The results of this study, which suggest absorption to be the main factor, are being published
separately (Wilkes \et\ 2005). Meanwhile, we report here the unusual X-ray spectrum of one AGN in the subset, \2M, a Type 1 QSO that
exhibits an absorbed power law continuum together with a soft excess. The suppression of the continuum below $\sim$1 keV allows the
form of the soft excess to be resolved with unusual clarity.

\section{Observation and data reduction} 

\2M (hereafter 2M23) is optically classified as a Type 1 QSO at a redshift of z = 0.199 (Cutri \et\ 2003). It was selected as a
`red' AGN from the 2 MASS catalogue (J-K$_{s}$=2.00$\pm$0.06) and found to have a linear broad band polarisation of $\sim$1 \%\ by
Smith et al (2002). It was observed by \xmm\ on 2003 July 3 (rev. 653) for 7885 s on target, with  X-ray data from the EPIC pn
(Str\"{u}der \et 2001) and MOS1 and MOS2 (Turner \et\ 2001) cameras providing moderate resolution spectra over the energy band
$\sim$0.2--10 keV. Each camera was set in full-frame mode, with the medium thickness filter to remove any optical/UV light from the
target source. The particle background was found to be high during parts of the observation, and those data have been excluded from
our  spectral analysis. Using the recommended maximum background rates of 1 s$^{-1}$ (pn camera) and 0.35 s$^{-1}$ (MOS camera) the
effective exposure times were thereby reduced to 4444 s (pn), 6321 (MOS1) and 6844 s (MOS2).

The X-ray data were screened with the XMM SAS v6.1 software and events corresponding to patterns 0-4 (single and double pixel
events) selected for the pn data and patterns 0-12 for MOS1 and MOS2. A low energy cut of 200 eV was  applied to all X-ray data and
known hot or bad pixels were removed. Source counts were obtained from a circular region of 45\arcs\ radius centred on the target
source, with the background being taken from a similar region offset from, but close to, the source. Individual EPIC spectra were
binned to a  minimum of 20 counts per bin to facilitate use of the $\chi^2$ minimalisation technique in spectral fitting, which was
based on the Xspec package (Arnaud 1996). All fits included absorption due to the line-of-sight Galactic column of
$N_{H}$=4.66$\times 10^{20}$ cm$^{-2}$. Errors are quoted at the 90\% confidence level ($\Delta \chi^{2}=2.7$ for one interesting
parameter).

\begin{figure}                                                          
\centering                                                              
\includegraphics[width=4.7cm, angle=270]{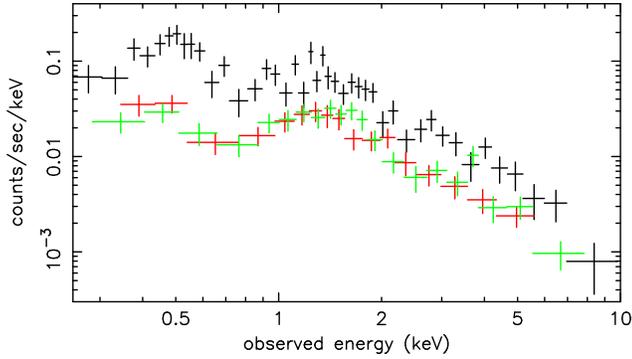}                     
\caption                                                                
{EPIC spectral data for \2M\ from the pn (black), MOS1 (red) and MOS2 (green) cameras, each showing evidence for
an absorption `trough' below $\sim$1 keV}      
\end{figure}

\section{The complex EPIC spectrum of \2M} 
\subsection{An absorbed power law}
Given the limited number of background-subtracted counts (844 pn and 790 MOS) spectral fitting was carried out on the integrated
data sets. Guided by the data (figure 1) we first fitted a power law above 2 keV, with a common spectral index, but untied
normalisations, for the 3 EPIC cameras, obtaining a  good fit ($\chi^2$=24 for 25 degrees of freedom) for a rather flat (hard)
photon index of $\Gamma$=1.65$\pm$0.2. Extrapolating  this fit to 0.2 keV showed the spectrum to be complex at low energies (figure
2) with evidence of additional (intrinsic) absorption and (possibly) a soft excess yielding a very poor overall fit
($\chi^2$=228/78).

\begin{figure}                                                          
\centering                                                              
\includegraphics[width=4.7cm, angle=270]{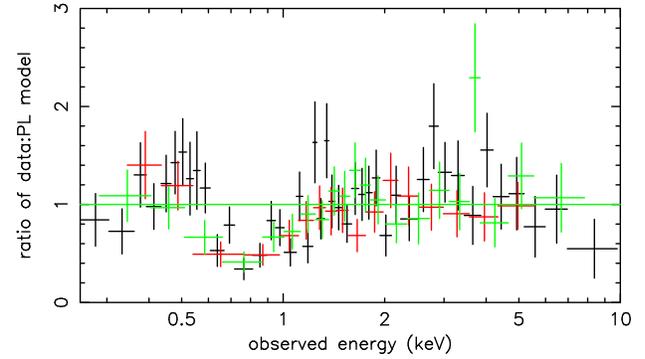}                     
\caption                                                                
{Ratio of the EPIC spectral data shown in figure 1 to a   
simple power law model fitted between 2-10 keV.  
The spectral structure indicates absorption below $\sim$1.5 keV and a `soft excess' near 0.5 keV}      
\end{figure}

To model the low energy absorption, we then added a photoionised absorber to the power law in Xspec, using a recent output (grid 18)
of the XSTAR code (Kallman and Bautista 2001). The ionisation parameter and column density were left as free parameters, but tied for the 3
data sets, while the key metal abundances of C--Fe were fixed at  their solar values. Grid 18 includes a turbulent velocity of 100
km s$^{-1}$. This addition yielded a much improved fit to the  combined pn and MOS data ($\chi^2$=89/76), with a column density
N$_{H}$=1.0$\pm$0.1$\times 10^{22}$ cm$^{-2}$ of moderately ionised gas.  The ionisation parameter $\xi$(=$L/nr^2$, where L is the
ionising luminosity irradiating matter of density n at a distance r) = 11$\pm$3 erg cm s$^{-1}$ was primarily constrained by the
spectral upturn, observed below $\sim$0.7 keV, which corresponds mainly to the absorption of ionised OVII in the model (figure 3).
Addition of the absorbing column resulted in a steepening of the power law index to $\Gamma$=2.05$\pm$0.14. Although the absorbed
power law fit was statistically acceptable a visual examination of the data:model residuals showed significant structure remaining
below$\sim$1 keV (figure 4).

\begin{figure}                                                          
\centering                                                              
\includegraphics[width=4.7cm, angle=270]{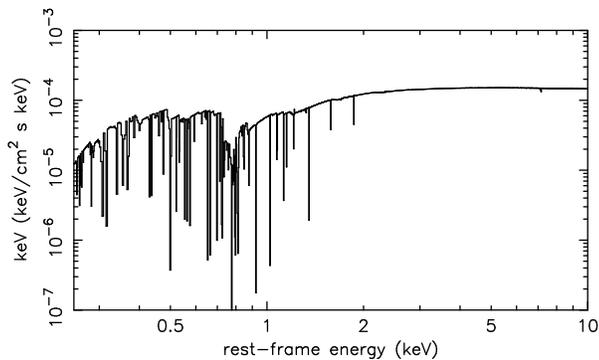}                                         
\caption                                                                
{Power law model fit to the \2M\ data attenuated by a column of `warm ' ionised gas
as described in Section 3.1. 
The energy axis is in the rest frame of \2M\ to aid 
visual identification of the spectral features}
\end{figure}

\begin{figure}                                                          
\centering                                                              
\includegraphics[width=4.7cm, angle=270]{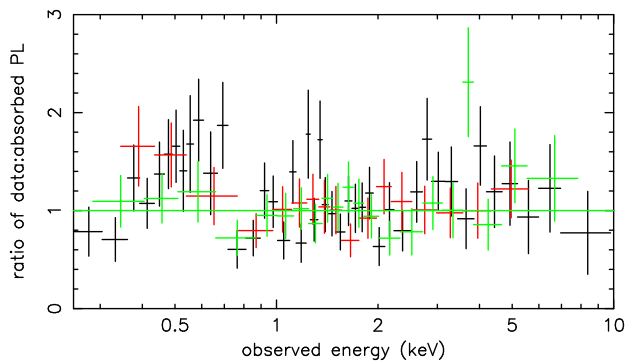}                     
\caption                                                                
{Ratio of the EPIC spectral data to the absorbed   
power law model of figure 3}      
\end{figure}

\subsection{A separate soft emission component}

An extended region of soft X-ray {\it emission} from outflowing photoionised gas is well established in Seyfert 2 galaxies (e.g.
Kinkhabwala \et\ 2002, Sako \et\ 2000), while {\it absorption} from `warm' gas is often seen in the X-ray spectra of Seyfert 1
galaxies (e.g. Kaspi \et\ 2002, Steenbrugge \et\ 2003). The residuals in figure 4 suggest both emission and absorption are affecting
the soft X-ray spectrum of 2M23. To test that we then added a soft X-ray emission component to the model for 2M23, lying outside the
cold absorbing matter affecting the power law continuum. 

To represent such a soft X-ray emission component we again used the XSTAR code, with the ionisation parameter and luminosity
(effectively the emission measure of the ionised gas) as  free parameters. Adding an emission  component to the absorbed power law
model in this way, with metal abundances again fixed at the solar values, yielded a further significant improvement to the 0.2-10
keV spectral fit ($\chi^2$=70/73). The standard F-test showed the improvement in the fit by adding the soft X-ray emission
spectrum was significant at 99.7\%. The ionisation parameter  of the emission component, essentially determining the energy profile
of the soft emission in the EPIC data, was $\xi$=18$\pm$11 erg cm s$^{-1}$. This more complex model is reproduced in figure 5.  The
dominant emission lines are seen to correspond to resonance transitions in He- and H-like ions of C, N, O and Ne, and Fe L (0.7-0.9
keV). Strong radiative  recombination continua of CVI, OVII and OVIII are also visible in the model spectrum. We deduce from this
XSTAR modelling that the $\it shape$ of the soft excess is well matched by the emission from a photoionised gas, the observed
luminosity (see below) corresponding  to an emission  measure  (EM=$\int$n$_{e}^{2}$dV) of order $2\times 10^{65}$ cm$^{-3}$.

\begin{figure}                                                          
\centering                                                              
\includegraphics[width=4.7cm, angle=270]{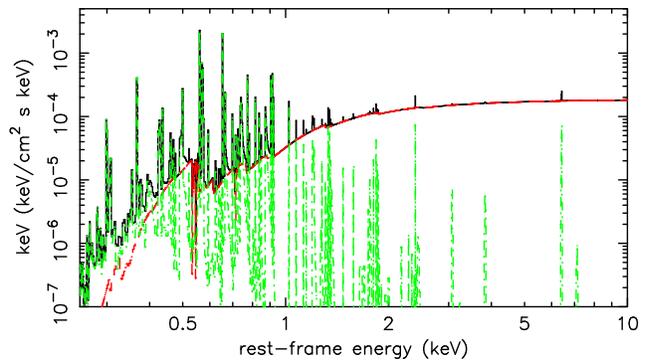}                                         
\caption                                                                
{Best-fit spectral model for \2M\ where the emission from ionised gas is added to the power law continuum, with the continuum now 
being
attenuated by relatively cold gas. 
The energy axis is in the rest frame of \2M\ to aid 
visual identification of the spectral features.
Details of this spectral fit are given in Section 3.2}
\end{figure}

A consequence of the additional soft X-ray emission was that the ionisation parameter of the absorbing column fell, to
$\xi$=0.1$\pm$0.1, now being determined mainly by the downward curvature in the data below $\sim$1 keV. The column density  of this
`cold' absorber also fell, to N$_{H}$=3$\pm$0.5$\times 10^{21}$ cm$^{-2}$, due to the higher relative opacity of the colder gas. 

The 0.2-2 keV luminosity in the `soft' emission component in this model (hereafter Model 1) was found to be $6.1\times 10^{42}$ erg
s$^{-1}$ (H$_{0}$=70 km s$^{-1}$ Mpc$^{-1}$). In comparison, the ${\it observed}$ 0.2-10 keV luminosity of the power law continuum
was $6.6\times 10^{43}$ erg s$^{-1}$ , increasing to $1\times 10^{44}$ erg s$^{-1}$ for a power law of $\Gamma$=1.9, correcting for
intrinsic absorption. We note the luminosity of the soft emission component is $\sim$18\% of the that removed from the power law
continuum by absorption, suggesting that the soft emission arises by re-emission from the same matter, a possibility discussed in
the next Section.

\subsection{Absorption and re-emission from the same ionised gas}

Given the implication in the above model of continuum energy being absorbed and then re-emitted in soft X-rays, it was important to
check whether this re-processing matter could be the same gas. To test that possibility, a further spectral fit was tried, tying
together the ionisation parameters of the absorbing and emitting matter in XSTAR. The fit was again acceptable ($\chi^2$=72/74),
for a tied ionisation parameter $\xi$=10$\pm$2 erg cm s$^{-1}$. This alternative `best fit' model had an unchanged  `normal' power
law with photon  index $\Gamma$$\sim$1.9, now absorbed  by a column N$_{H}$$\sim$1$\times 10^{22}$  cm$^{-2}$ of warm (moderately
ionised) gas. Figure 6 illustrates this model (hereafter Model 2).  

\begin{figure}                                                          
\centering                                                              
\includegraphics[width=4.7cm, angle=270]{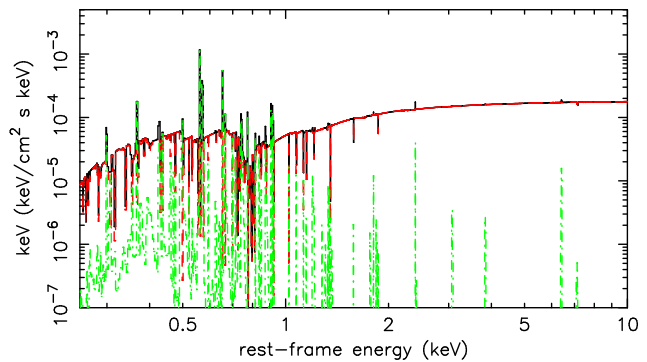}                                         
\caption                                                                
{Alternative best-fit spectral model for \2M\ where the ionisation parameters of the absorbing and emitting gas are tied.
The energy axis is in the rest frame of \2M\ to aid 
visual identification of the spectral features.
Details of this spectral fit are given in Section 3.3}
\end{figure}

Although not quite such a good fit as Model 1, the common warm absorber/re-emitter model for the \xmm\ EPIC  spectrum of 2M23 has the
attraction of involving one less physical component. Visual comparison of figures 5 and 6 shows the  soft X-ray emission to be
somewhat weaker in Model 2. Specifically, the 0.2-2 keV luminosity in the `soft' emission component of Model 2 fell to $2.5\times 10^{42}$
erg s$^{-1}$, while the ${\it observed}$ 0.2-10 keV luminosity of the power law continuum of 2M23 was $7\times 10^{43}$ erg s$^{-1}$,
or $1\times 10^{44}$ erg s$^{-1}$ correcting for intrinsic absorption. In Model 2 the luminosity of the soft emission component is 
then $\sim$8\% of that removed from the power law continuum by absorption, a factor $\sim$2 less than in Model 1.

\section{Discussion}
The \xmm\ spectrum of 2M23 has a typical Type 1 AGN power law continuum slope ($\Gamma$$\sim$1.9) when allowance is made for
absorption. The nature of the line-of-sight column is unclear from our fitting, the low energy opacity (larger than usual for a Type
1 QSO) being modelled by a column N$_{H}$$\sim$3$\times 10^{21}$  cm$^{-2}$ of `cold' gas, or N$_{H}$$\sim$1$\times 10^{22}$  
cm$^{-2}$ of moderately ionised or `warm' gas. Importantly, the attenuation of the power law component below $\sim$1 keV allows the
spectral shape of the `soft excess' to be  resolved and quantified.

In the first spectral fit (Model 1) the X-ray absorbing gas is cold and embedded dust could also be responsible for the optical 
reddening seen in this 2MASS QSO. While it is possible that the BLR has a flattened geometry and therefore suffers less absorption
than the (centrally located) continuum X-ray source, the Type 1 optical classification of 2M23 suggests such `cold' matter lies close
to the continuum source. If so, it must be relatively dense to survive.

We note the growing evidence for `variable' cold absorbing matter in AGN (eg Risaliti \et\ 2002). On the large scale this is seen in
the tendency for some type 2 AGN switching between Compton-thick and Compton-thin X-ray absorption (eg Guainazzi \et\ 2004). Again, a
substantial column density of near-neutral matter was seen to shrink as the continuum X-ray brightness increased strongly in the
luminous type 1 Seyfert 1H0419-577 (Pounds \et\ 2004a). It is interesting to speculate that the `cold' X-ray absorption in Type 1 AGN
might be associated with the accretion process and related, for example, to the accreting clouds envisaged by Guilbert and
Rees (1988) in their original paper predicting the subsequent discovery of X-ray `reflection' (Pounds \et\ 1990). Alternatively it
may lie in dense structures at the  base of an outflow or wind.

In the alternative spectral fit to 2M23 (Model 2) the absorbing gas has the same ionisation parameter as that of the soft X-ray
emission. However, in the context of the `red' near-infrared colour of 2M23, it is noteworthy that the ionisation parameter
($\xi$$\sim$10) is lower than commonly found in Type 1 AGN. In this case the combination of a relatively low
ionisation parameter and unusually high column density, N$_{H}$$\sim$1$\times 10^{22}$  cm$^{-2}$, may be causally linked with the 
designation of 2M23 as a 2MASS QSO. 

The main result from our spectral modelling of 2M23 is that the soft excess - revealed with unusual clarity due to the effective
removal of the soft power law component - has a spectral form well matched by the blended line emission from an ionised gas. A
similar finding was reported in the low flux state spectrum of the Seyfert 1 galaxy NGC 4051 (Pounds \et 2004b), and proposed as the
origin of a quasi-constant soft X-ray emission component in the luminous Seyfert 1 galaxy 1H0419-577 (Pounds \et\ 2004a). It was
speculated in those papers that such an emission component might be a stronger analogue of the  extended soft X-ray emission seen in
Seyfert 2 galaxies such as NGC 1068 (Brinkman \et\ 2002, Kinkhabwala \et\ 2002) and Mkn 3 (Sako \et\ 2000, Pounds and Page 2005). In
both those bright, nearby Seyfert 2 galaxies the high resolution soft X-ray spectrum is dominated by line emission from photoionised/
photoexcited gas, with the scattered continuum being weak. The shape of the soft emission in 2M23 suggests this is again the case. 

To quantify the comparative strength of the soft X-ray emission in these Type 1 AGN with an archetypal Type 2 AGN, we list in Table 1
their observed and derived soft X-ray luminosities, together with the intrinsic 2-10 keV luminosities as a proxy for the relative
ionising fluxes.  From this admittedly small sample it is seen that the luminosity ratio of the soft X-ray emission to power law is
up to an order of  magnitude higher in the Type 1  AGN - including 2M23, compared with Mkn3. 

In considering a common origin for the soft X-ray emission, we surmise that the ionised outflow seen in a Type 2 AGN is cut off  from
direct view at some minimum radius $r$ by the same obscuring matter that hides the BLR, while in a Type 1 AGN the full extent of the
outflow is visible. With the simplest assumption of a constant velocity, radial outflow, mass conservation yields a {\it total}
emission measure of ionised gas which increases as $r^{-1}$.  That would require the ionised gas to extend inward to 0.1$r_{cut}$,
where $r_{cut}$ is the minimum radius observed directly in Type 2 AGN. Taking $r_{cut}$ as intermediate between the BLR and NLR, and
a discriminating Keplerian velocity of $2\times10^{3}$ km s$^{-1}$, $r_{cut}$ is then $\sim$$ 10^{4}$$r_{g}$, where $r_{g}$ is the
gravitational radius. To obtain a 10-fold increase in the integrated soft X-ray emission then implies the ionised flow in 2M23 to
extend inward to $\sim$$10^{3}$$r_{g}$. Is that consistent with observation?

\begin{table*}
\centering
\caption{The soft X-ray and intrinsic 2-10 keV luminosities of \2M\ and two other Type 1 AGN, compared with those of the archetypal
Type 2 AGN, Mkn3.  
The 2-10 keV luminosities are used as a proxy for the ionising flux irradiating the soft X-ray emission region in each case}

\begin{tabular}{@{}lccccccc@{}}
\hline
Galaxy & optical type & L$_{2-10}$(ergs s$^{-1}$) & L$_{SX}$(ergs s$^{-1}$) &  ratio (L$_{SX}$/L$_{2-10}$)  & reference \\

\hline
NGC 4051 & Seyfert 1 & $3.5\times 10^{41}$ & $3.5\times 10^{40}$ & 10\% & Pounds \et\ 2004b \\
1H0419-577 & Seyfert 1 & $2.7\times 10^{44}$ & $2.9\times 10^{43}$ & 11\% & Pounds \et\ 2004a \\
2M234449 & QSO 1 & $5.3\times 10^{43}$ & $6.1\times 10^{42}$ & 11\% & Model 1, this paper \\
2M234449 & QSO 1 & $5.3\times 10^{43}$ & $2.5\times 10^{42}$ & 5\% & Model 2, this paper \\
Mkn 3 & Seyfert 2 & $2\times 10^{43}$ & $2.5\times 10^{41}$ & 1.2\% & Pounds and Page 2005 \\

\hline
\end{tabular}
\end{table*}

The escape velocity of gas at $\sim$$10^{3}$$r_{g}$ is $\sim$$10^{4}$ km s$^{-1}$. Turbulence in such gas would result in very broad 
wings to the X-ray absorption lines, apparently at odds with reported \xmm\ and \chandra\ grating spectra of BLAGN  which are an
order of magnitude narrower (e.g. Steenbrugge \et\ 2003, Kaspi \et\ 2002). However, those high dispersion instruments are notably
insensitive to broad features and broad wings could well remain undetected. A similar - but  more extreme - scenario was
envisaged by Gierlinski and Done (1994) in suggesting low energy absorption in highly turbulent matter as an explanation of the
common form of the soft X-ray spectra in many AGN.

In our proposed interpretation of the soft excess in 2M23 we would expect the energy absorbed from the power law continuum by
line-of-sight matter to be related to the luminosity of an optically thin ionised gas by the the covering factor. The values we
find,  of $\sim$8-17\%, are consistent with the typical solid angle of $\sim$1 sr {\it observed} in the biconical outflows of Type 2
AGN.

\section{Summary} 
The \xmm\ EPIC spectrum of the Type 1 QSO \2M\ is unusual in showing low energy absorption at a level that allows a soft X-ray
emission component to be resolved. Modelling of the absorption and emission with the XSTAR code allows the form and strength of the
soft X-ray emission to be determined and shown to be consistent with the integrated emission from a warm  photoionised gas. A
plausibility argument is given which suggests this enhanced soft X-ray emission (which should be a common feature in Type 1 AGN)
arises from an inward extension of the outflow responsible for the soft X-ray emission in Seyfert 2 galaxies. This is important in
the context of the energy content, emission measure and origin of such outflows, since current high resolution spectra are not
well-matched to the detection of broad spectral features that might arise in turbulent gas at the smaller radii implied by our
analysis.    

\section{Acknowledgments} The results reported here are based on observations obtained with \xmm, an ESA science
mission with instruments and contributions directly funded by ESA Member States and the USA (NASA). The authors wish
to thank the SOC and SSC teams for organising the \xmm\ observations  and initial data reduction. KAP is pleased to
acknowledge a Leverhulme Trust Emeritus Fellowship and KLP funding from PPARC.  BJW is grateful for the financial support of \xmm\ GO grant:
NNG04GD27G.

\end{document}